\documentclass[12pt]{article}
\usepackage{fullpage}
\usepackage{overcite}
\usepackage{epsfig}
\usepackage{subfigure}

\begin{document}

\noindent
{\LARGE \bf \sf A liquid-solid critical point in a simple
monatomic system}

\vskip 0.5cm

\noindent
M\aa ns Elenius$^1$ and Mikhail Dzugutov$^2$

\vskip 0.5cm

\noindent
{\it $^1$Department of Numerical Analysis and ~$^2$Dept. of
Materials Science and Engineering, Royal Institute of Technology,
SE-100 44 Stockholm, Sweden}

\vskip 0.5cm

\noindent
 
\abstract{It is commonly believed that the transition line
  separating a liquid and a solid cannot be interrupted by a
  critical point. This opinion is based on the traditional
  symmetry argument that an isotropic liquid cannot be
  continuously transformed into a crystal with a discrete
  rotational and translational symmetry. We present here a
  molecular-dynamics simulation of a simple monatomic system that
  demonstrates a liquid-solid spinodal terminating at a critical
  point. We show that, in the critical region, the isotropic
  liquid continuously transforms into a phase with a mesoscopic
  order similar to that of the smectic liquid crystals. We argue
  that the existence of both the spinodal and the critical point
  can be explained by the close structural proximity between the
  mesophase and the crystal. This indicates a possibility of
  finding a similar thermodynamic behaviour in gelating colloids,
  liquid crystals and polymers.}

\newpage

\section{Introduction}
In many systems, a domain of spinodal instability can be observed
where a liquid phase coexists in equilibrium with a gas phase. The
spinodal terminates at a critical point \cite{landau} which manifests
a possibility of a continuous transformation between the two phases.
Whether a single-component system can exhibit a spinodal where a
liquid and a crystalline solid phase coexist in equilibrium, and a
respective critical point, still remains an unanswered question of a
significant conceptual interest. A liquid-solid critical point has
never been observed hitherto, and its existence is commonly regarded
impossible based on the argument that a continuous transformation from
an isotropic liquid to a crystal with a discrete symmetry is
inconceivable \cite{landau}.

Here, we report a molecular dynamics simulation of a simple
monatomic system that demonstrates under cooling a first-order
transition from a liquid phase into a spinodal domain where the
liquid coexists in equilibrium with a crystalline phase. Within
the spinodal, the relative volume occupied by the crystalline
phase shrinks with density reduction, vanishing in the
low-density limit where the spinodal is terminated by a critical
point. It is demonstrated that, upon cooling to the vicinity of
the critical point, the isotropic liquid continuously transforms
into a low-density phase with a mesoscopic order similar to that
of smectic liquid crystals \cite{anisimov}. Furthermore, we
observed a close structural proximity between the mesophase and
the crystal which extends much beyond the local order of nearest
neighbours. We argue that this structural peculiarity of the
mesophase can account for the existence of both the spinodal and
the critical point. It also makes it possible to conjecture that
a liquid-solid critical point like the one we observe here might
exist in  liquid crystals, polymers and in gelating
colloidal systems.

The following crystallization scenario suggested by the classical
nucleation theory is commonly accepted. In a metastable supercooled
liquid, nuclei of the lower free energy crystalline phase are being
spontaneously created by virtue of thermal fluctuations. Because the
two phases are symmetry-unrelated, appearance of a nucleus involves
creation of an interface. The free-energy cost of the interface blocks
the growth of the nucleus unless the latter exceeds a critical size.
In this scenario, the entire volume of a metastable supercooled liquid
rapidly crystallizes as soon as a critical nucleus has been created.
The existence of a spinodal instability domain where no single phase
can be in a stable or metastable equilibrium implies that the
interfacial free-energy barrier vanishes.

In general, there are two independent order parameters involved
in a first-order phase transition: density and the local
structural symmetry. The former controls fluid-fluid phase
transitions where a spinodal instability domain is commonly
observed, whereas the latter plays the key role in
crystallization of dense liquids. However, it was pointed out
that density fluctuations seen in the vicinity of a metastable
liquid-liquid spinodal with a critical point may cause a dramatic
increase of the nucleation rate, indicating a reduction of the
interfacial barrier for crystallization \cite{frenkel1, olmsted,
oxtoby}. The possibility of a liquid-solid spinodal was also
inferred from the crystallization anomalies in simple liquids
under deep supercooling \cite{klein, frenkel2, parrinello} and
phase transformations in atomic clusters \cite{ berry,
walesberry, walesdoye, saika}. It was also predicted, based on
analytic arguments, that a fluid-solid spinodal region may appear
in the system of hard spheres in higher dimensions
\cite{kleinfisch}

A possible microscopic mechanism for the density-driven spinodal
domain in the liquid-solid transition of a simple
single-component system was conjectured by Debenedetti et al.
\cite{debenedetti} based on a lattice model. A pair potential,
attractive at the nearest-neighbour distance and repulsive at the
second-nearest neighbour distance, was shown to induce
low-density open structures, associated with the liquid phase,
which collapse into a dense solid phase under pressure. Due to a
significant difference in density, the two phases can coexist
within a spinodal domain of infinite compressibility.

\section{Simulation details}
In the present simulation study, we explore the type of the pair
potential suggested in the above conjecture in a molecular
dynamics simulation of a single-component system. The pair
potential we exploit \cite{doye} (denoted as Z2 in that
reference) has been found to induce a pronounced icosahedral
local ordering of the nearest neighbours due to the design of its
short-range attraction, while the repulsion incorporated in its
longer-range part inhibits bulk packing of icosahedra. In this
simulation, we used a system of 3456 identical particles in the
NVT ensemble. All results here are presented in reduced
simulation units following from the definition of the pair
potential \cite{doye}.

\section{Results}
The phase behaviour of the simulated system was explored within a wide
range of densities by cooling it at constant density $\rho$ from the
high-temperature isotropic liquid state. The cooling was performed
stepwise, with comprehensive equilibration at each temperature
point. The isochoric variation of enthalpy $H$ as a function of
temperature $T$ is shown in Fig. 1a. For each isochore where
$\rho>0.32$ we observe a discontinuity in $H(T)$ indicating a
first-order freezing transition into a low-temperature state. The
first-order character of this transition is confirmed by distinct
hysteresis loops in the $H(T)$ isochores that are observed when the
system is heated from the low-temperature state towards the
high-temperature liquid.

Both the freezing and melting temperatures derived from the isochoric
enthalpy variations are shown in the $\rho - T$ phase diagram, Fig.
1b. A distinctive anomaly of the observed phase transition is that the
melting line and the freezing line confining the metastability domain
converge at $\rho = 0.33$ and $T=0.31$. The first-order phase
transition apparently disappears at densities below $\rho = 0.33$.
This is confirmed in Fig. 1c that shows the density dependence of the
latent heat $\Delta H$ of this transition. That quantity reduces in an
apparently linear manner with the reduction of density, eventually
vanishing at the indicated density point. Indeed, no trace of
discontinuity or hysteresis can be discerned in the isochoric
variation of enthalpy within the relevant temperature interval at
$\rho = 0.32$ presented in Fig. 1a. These observations lead to the
conclusion that at $\rho = 0.33$ and $T=0.31$ we observe a critical
point which terminates the first-order transition line.

Another interesting anomaly of this phase transition can be observed
in the behaviour of the isothermal compressibility $\chi_T$. It is
related to the low-$Q$ limit of the structure factor $S(Q)$ as $\chi_T
=\rho T S(0)$. The structure factor is defined as $S(Q) = \langle
\rho({\bf Q})\rho(-{\bf Q})\rangle$, where $Q=|{\bf Q}|$ and
$\rho({\bf Q})$ is the Fourier transform of the system's density:
\begin{equation}
  \rho({\bf Q}) = \frac{1}{N} \sum_{k=1}^{N}\exp (-i{\bf Q r}_k) 
\end{equation}
$N$ being the number of particles and ${\bf r}_k$ the real space
location of particle $k$. Fig. 1d shows the density dependence of
$S(0)$ in the low-temperature state along the freezing line. Within
the density interval above the estimated critical density $S(0)$
evidently demonstrates a pronounced tendency for divergence. The
observed behaviour of the isothermal compressibility indicates the
system's large-scale decomposition into domains of different density,
a characteristic signature of spinodal instability. We stress that
this result was obtained after a thorough equilibration of the
low-temperature state and, therefore, cannot be regarded as a
transient effect of the phase transition.

In order to gain an insight into the microscopic mechanism of these
thermodynamic anomalies, we now analyse the structural aspects of the
observed phase transformation by directly inspecting the atomic
configurations of the low-temperature state. We note that, to get rid
of the thermal noise, all the configurations investigated here were
subjected to steepest-descent minimization of potential energy.

Fig. 2 depicts a set of configurations representing the system's
states equilibrated at $T$ immediately below the freezing
transition, at different densities. At $\rho = 0.5$, the
transition evidently produces a structure characterised by a
well-defined global periodic order. This is a low-density
crystalline phase with a local order apparently dominated by
tetrahedral close packing. The central element of its unit cell
is a cubically-symmetric configuration composed of eight
icosahedra which are kept apart by low-dimensional arrangements
of tetrahedra. In a similar manner, low-dimensional tetrahedral
bridges link the described cubes into a periodic structure. To
the best of our knowledge, this crystalline structure has not
been reported so far. A detailed crystallographic description of
this phase will be reported elsewhere.

Other atomic configurations shown in Fig. 2 b-d represent the density
domain of diverging isothermal compressibility that we identified as
spinodal. Within each of these configurations, we are able to discern
a distinct compact domain of the same crystalline structure as the one
observed at $\rho = 0.5$. These crystalline domains are marked by red
in the plots. The rest of the volume of the MD simulation cell outside
the crystalline domains is occupied by an apparently different
structure that is visibly less dense and doesn't exhibit any
immediately detectable global order. Its local order, however, appears
to be quite similar to that of the crystalline structure, being
dominated by the same kind of tetrahedral close packing with a high
number of icosahedra linked in a low-dimensional manner. The plots in
Fig. 2 also demonstrate that the relative volume occupied by the
crystalline phase shrinks approximately linearly with reduction of
density, eventually vanishing at the critical point. This observation
is evidently consistent with the respective variation of the latent
heat shown in Fig. 1c. It demonstrates that the liquid phase can
be continuously transformed into a crystal via the critical point.

We now analyse the transformation of the liquid phase under
cooling.  First, we consider the isochore $\rho=0.32$, i.e. below
the critical density. Temperature variations of the specific heat
$C_V$ at that density, shown in Fig. 3a, displays a pronounced
maximum at $T_m=0.45$. The density dependence of $T_m$ is shown
in the phase diagram, Fig. 1b. The line of $C_V$ maximum extends
to higher densities until it eventually encounters the freezing
transition line.  The appearance of excess specific heat under
cooling is an indicator of the development of bonded structures,
and a common precursor of percolation \cite{angell99,
sciortino1}. Indeed, a steep increase in the number of icosahedra
can be observed below $T_m$, Fig. 3b. This leads to the
development of extended clusters and a rapid growth of the
correlation length, as it is demonstrated by the
temperature-dependence of the maximum cluster size,
Fig. 3b. These clusters are extended low-dimensional structures
composed of interconnected icosahedra and face sharing
tetrahedra, like the one exemplified in Fig. 3c. Upon further
cooling, the maximum cluster size diverges, and at $T=0.28$, a
continuous tetrahedrally packed network is formed that includes
approximately 90 \% of all the atoms. An example of that network
is shown in Fig. 2e. Another interesting structural feature
developed in this liquid state under cooling is a low-$Q$
pre-peak of $S(Q)$, Fig. 3d. This anomaly, as well as the maximum
of $C_V$, are commonly observed in the context of gelation in
colloidal systems \cite{sciortino2}.

Fig. 3e presents the ratio of the number of icosahedra in the
high temperature state to the respective number in the low
temperature state within the metastability domain, as a function
of density. It indicates that, as the density is reduced, the
difference in the local order between the two states shrinks,
vanishing at the critical point.  A considerable degree of local
structural similarity between the liquid in the critical-point
region and the crystal is exemplified by the ring-like
arrangement of icosahedra depicted in Fig. 2e. A trend for the
formation of similar structural arrangement, although in a more
defective rudimentary form, can be detected in the
higher-temperature cluster, Fig. 3c. Moreover, the liquid
exhibits an instantaneous mesoscopic-scale periodicity consistent
with that of the crystal. Fig. 4a shows the intensity pattern of
the liquid's $S({\bf Q})$ on a sphere in the ${\bf Q}$-space, with
a radius equal to $Q_{pp}$, the position of the low-$Q$ pre-peak
of $S(Q)$; Fig. 4b shows the respective pattern exhibited by the
crystal on the ${\bf Q}$-sphere of the same radius. The liquid
diffraction pattern shows a sizable degree of periodicity along
two perpendicular axis, consistent with the respective
periodicity pattern in the crystal. The observed periodic order
in the liquid is however limited in both time and space. Its
dissipation in time is demonstrated in Fig. 4c where we present
the decay of the respective component of the density-density
correllator $F(Q_{pp},t) = \langle \rho( Q_{pp},0)\rho(-
Q_{pp},t)\rangle$, for a set of temperatures.  The
density-correlation pattern manifested by the pre-peak of $S(Q)$
evidently dissipates. The relaxation time exhibits a strong
temperature-dependence, and exceeds the timescales accessible in
our simulation for temperatures below $T=0.28$.

\section{Discussion}

The liquid phase formed by cooling from the high-temperature
liquid at $\rho=0.32$ can be classified as a mesophase similar to
the smectic phases \cite{anisimov} observed in polymeric
liquids. We also mention that a liquid-crystal smectic phase
interposed between the isotropic liquid and the crystalline solid
was discussed \cite{nelson} as a possible scenario for the bond
ordering in the supercooled liquid state.  Both melting of such a
phase and its transformation to a crystal were conjectured to be
continuous. The present result supports that conjecture. Another
possible candidate for the observation of a liquid-solid critical
point may be polymers where a liquid-solid spinodal was recently
observed \cite{gee}.

Based on these observations, we can understand the existence of a
spinodal domain where the mesophase coexists with the crystal,
and the critical point considering two factors. First, there is a
significant difference in density between the two coexisting
phases. The characteristic low-density open structure of the
mesophase is induced by the pair potential which enforces a
strong tendency for icosahedral ordering of the nearest
neighbours while keeping the icosahedra apart due to the
repulsion incorporated in its long-range part. Second, the
structural proximity between the mesophase and the crystal that
extends well beyond the nearest-neighbours' configurations can be
expected to greatly reduce the free-energy cost for the interface
separating the two structural domains.  These two factors render
the system unstable with respect to density fluctuations of a
large spatial scale, making possible both an equilibrium
coexistence of the two phases within the spinodal domain and the
existence of the critical point at its low-density limit.  In
this way, the present observation can be compared with the
coexistence of structurally similar liquid phases discriminated
by density, and a respective critical point observed in a
simulation of a one-component system using a pair potential with
a long-range repulsion \cite{buldyrev}.

The structural similarity between the mesoscopic liquid state formed
by the present system under cooling at sub-critical density and
gel-forming colloidal systems that we have already mentioned deserves
an additional comment. It was found \cite{sciortino3} that the
appearance of spanning clusters involved in gelation in low-density
systems of colloidal particles with short-range interaction can be
interpreted in terms of density fluctuations caused by the existence
of a liquid-gas spinodal instability. A possible relation between that
instability and the higher-density liquid-solid critical point we
report here looks quite intriguing, and deserves to be investigated in
future studies. In particular, a question of significant interest is
whether the tetrahedrally-packed colloidal clusters demonstrating
gelation at low densities \cite{sciortino1, sciortino2} would at
higher densities display coexistence between a disordered phase and a
crystal, and a respective liquid-solid critical point like the one we
report here.

\section{Conclusions}

In summary, the molecular dynamics simulation presented here
demonstrates a novel kind of phase behaviour in a simple
one-component system.  Its phase diagram is found to include a
spinodal domain separating a stable liquid phase and a
crystalline solid phase where the two phases coexist in
equilibrium.  Moreover, it is demonstrated that the spinodal
terminates at a critical point. The existence of a liquid-solid
critical point implies that an isotropic liquid can be
transformed into a crystal without crossing the line of
first-order phase transition. We argue that the existence of
both the spinodal and the critical point can be explained by the
structural transformation of the liquid phase.  Upon cooling
towards the critical point, the liquid is found to develop a
mesoscopic order, reminiscent of that in smectic liquid crystals,
which demonstrates a strong structural similarity with the
crystalline phase. We conjecture that the observed phase
behaviour can be expected in other systems possessing a similar
kind of mesoscopic order, including liquid crystals, polymers and
gelating colloidal systems.

\section*{Acknowledgements}
We would like to thank Prof. F. Sciortino, Prof. D. Frenkel and Prof.
D.~J. Wales for reading the manuscript and very useful comments.

\newpage
Captions\\
\\
Figure 1 \\
Liquid-solid phase transformation. {\bf a}. Isochoric
temperature variations of the enthalpy $H$ at indicated densities.
Blue dots and line: cooling; red dots and line: heating. {\bf b}.
Density-temperature phase diagram of the simulated system. Blue
symbols and line: the freezing transition; triangles: crystallization;
circles: spinodal. Green symbols and line: melting transition. Red
line and symbols denote maximum of $C_V$ in the liquid phase. MP and
CP respectively abbreviate mesophase and critical point. {\bf c}.
Density-dependence of the latent heat $\Delta H$ for the freezing
transition. {\bf d}. Density-dependence of the estimated small-$Q$
limit of the structure factor $S(Q)$ at the freezing line.
\\
\\
Figure 2\\
  Instantaneous configurations of the low temperature
state. {\bf a}. $\rho=0.5$, $T=0.48$. {\bf b}. $\rho=0.4$, $T=0.41$.
{\bf c}. $\rho=0.36$, $T=0.36$. {\bf d}. $\rho=0.34$, $T=0.3125$. {\bf
e}. $\rho=0.32$, $T=0.28$. The cubic arrangement of 8 icosahedra shown
separately in {\bf a} represents the central element of the
crystalline phase. In {\bf b-d}, red colour marks crystalline domains
coexisting with the liquid phase. Separate plots present the same
domains in a different projection. In {\bf e}, the configuration
domain marked by red, also shown in a separate plot, is an arrangement
of four icosahedra, with some defects, representing a rudiment of the
crystalline structure.
\newpage

Figure 3\\
    Evolution of the liquid phase under cooling. {\bf
a}.  Temperature variation of the liquid specific heat $C_V$
along the isochore $\rho=0.32$. {\bf b}. Temperature variation,
along the isochore $\rho=0.32$, of the number of icosahedra $N_i$
(black) and the number of atoms $N_c$ in the maximum-size cluster
composed of face sharing tetrahedra, scaled by the total number
of atoms in the system $N$ (blue). {\bf c}. A maximum-size
tetrahedral cluster observed at $\rho=0.32, T=0.34$. {\bf
d}. Temperature variation of the structure factor $S(Q)$. Black,
$T=1.10$; green, $T=0.40$; red, $T=0.28$. {\bf e}. The ratio of
the number of icosahedra in the high temperature state, $N_{HT}$,
to the respective number in the low temperature state, $N_{LT}$,
within the metastability domain, as a function of density.
\\
\\
Figure 4
\\
Mesoscopic-scale anisotropy in the low-temperature liquid
phase. {\bf a}. Intensity of the structure factor $S(\bf{Q})$ on a
${\bf Q}$-space sphere of the radius $Q_{pp}$, the position of the
low-$Q$ pre-peak of $S(Q)$ for a liquid configuration at $\rho=0.32$,
$T=0.28$. {\bf b}. The same plot as Fig. 4a but for the crystalline
phase at $\rho=0.5$, $T=0.48$. {\bf c}. The decay of the
density-density correllator $F(Q,t)$ at $\rho=0.32$. Red lines: $Q$
corresponding to the position of the main peak of $S(Q)$. Blue lines:
$Q=Q_{pp}$. From left to right, $T=0.30$, $0.29$, $0.28$, $0.27$.

\newpage

\begin{figure}
  \includegraphics[width=16.cm]{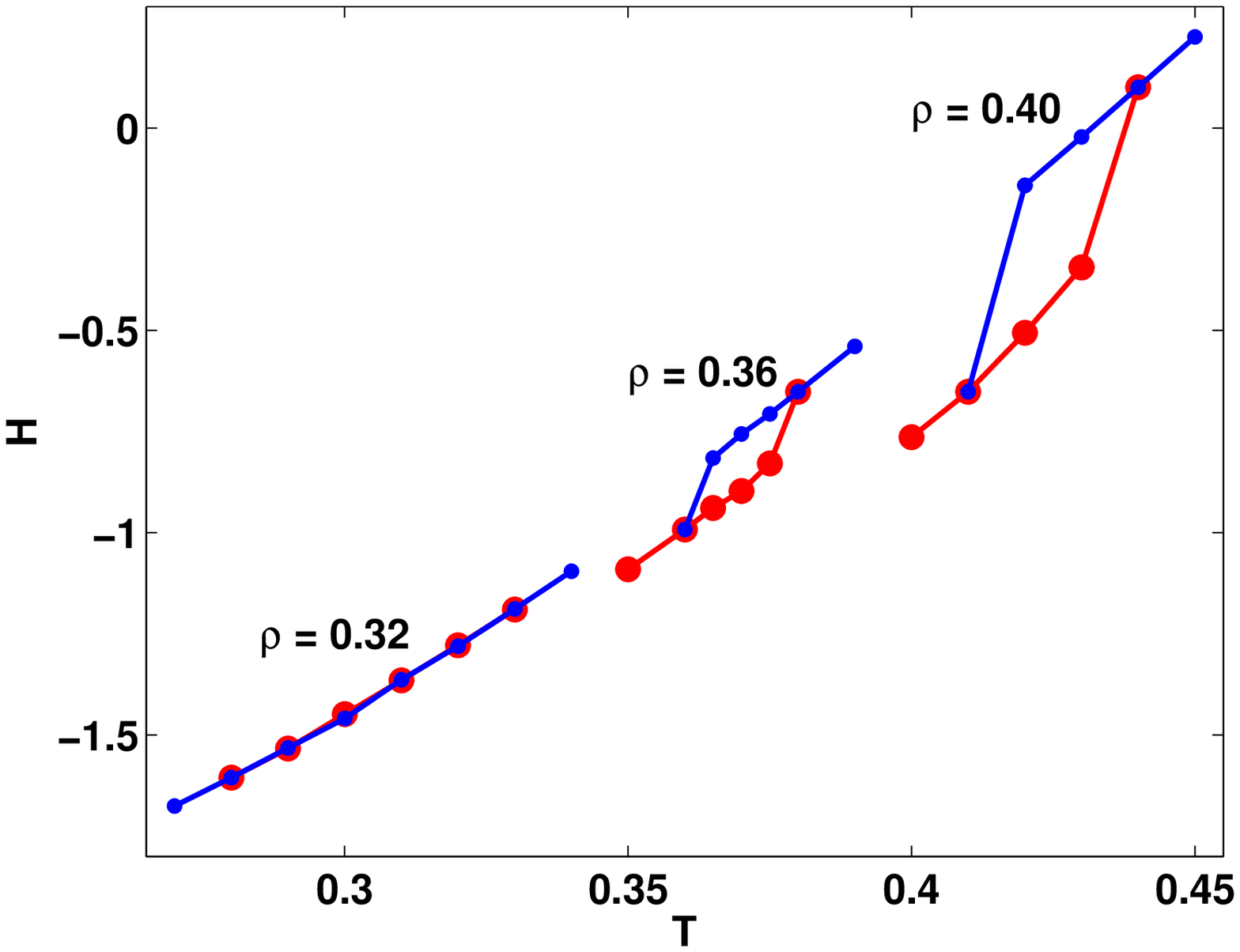}
\end{figure}

1a

\newpage
\begin{figure}
  \includegraphics[width=16.cm]{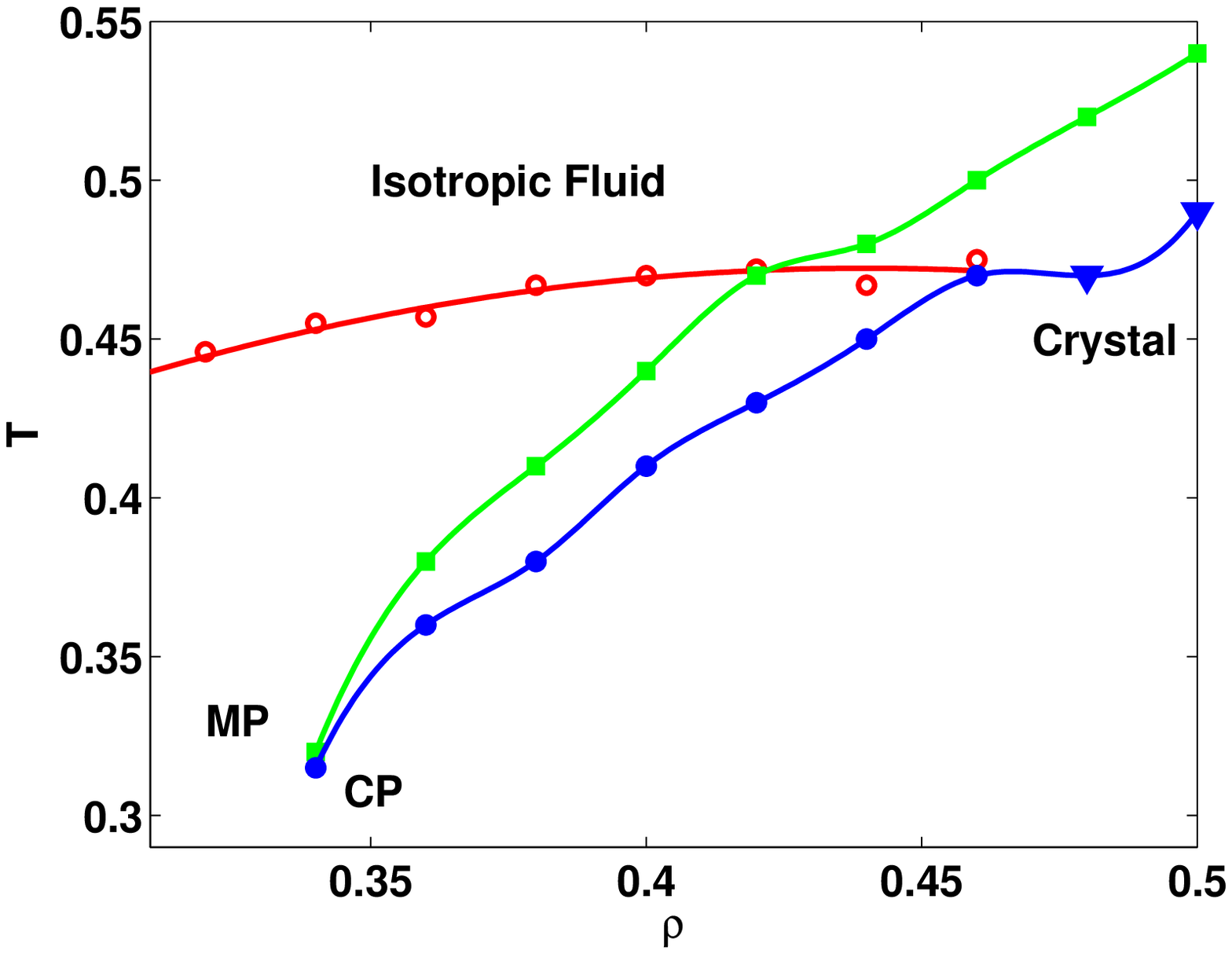}
\end{figure}

1b

\newpage
\begin{figure}
  \includegraphics[width=16.cm]{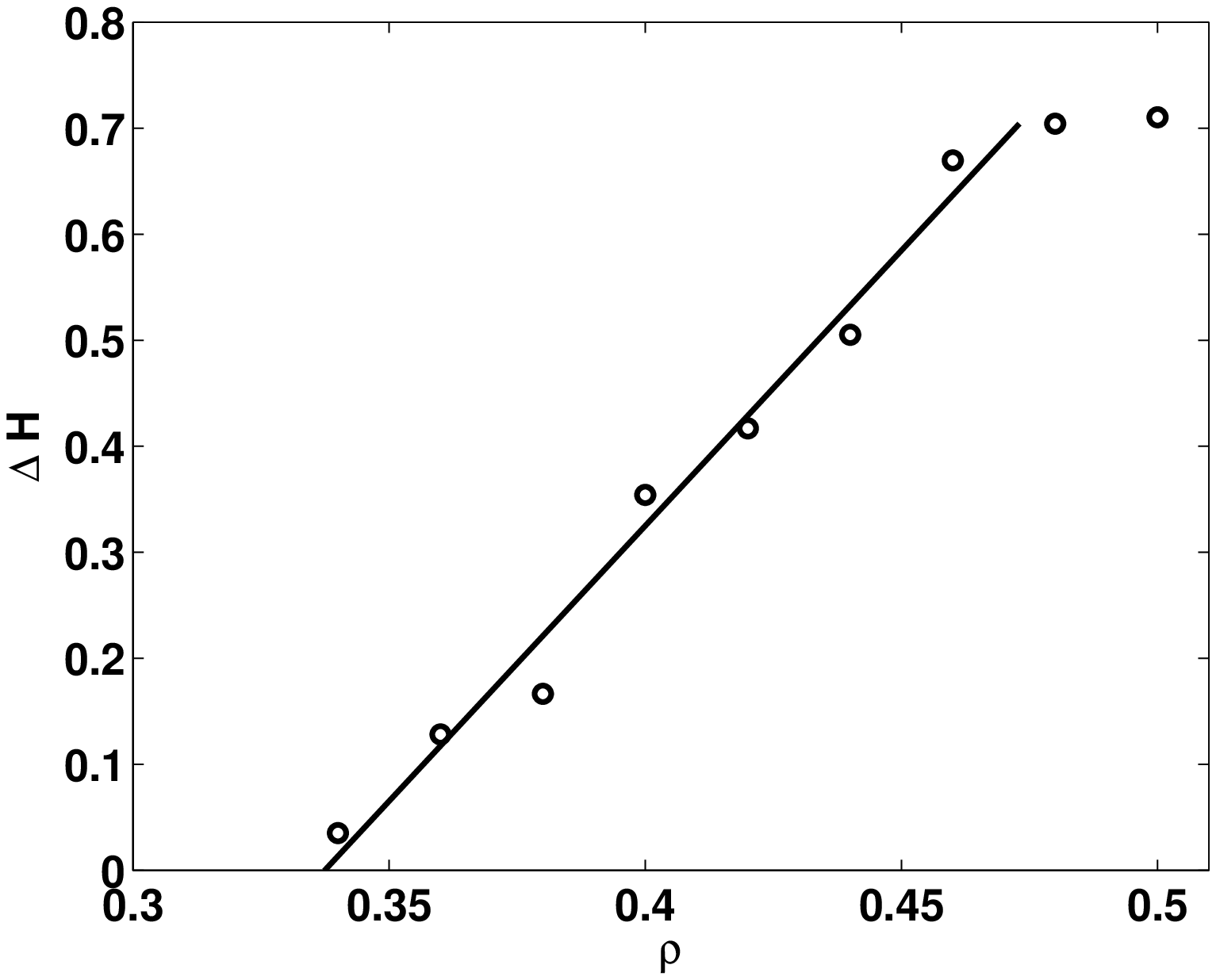}
\end{figure}

1c

\newpage
\begin{figure}
  \includegraphics[width=16.cm]{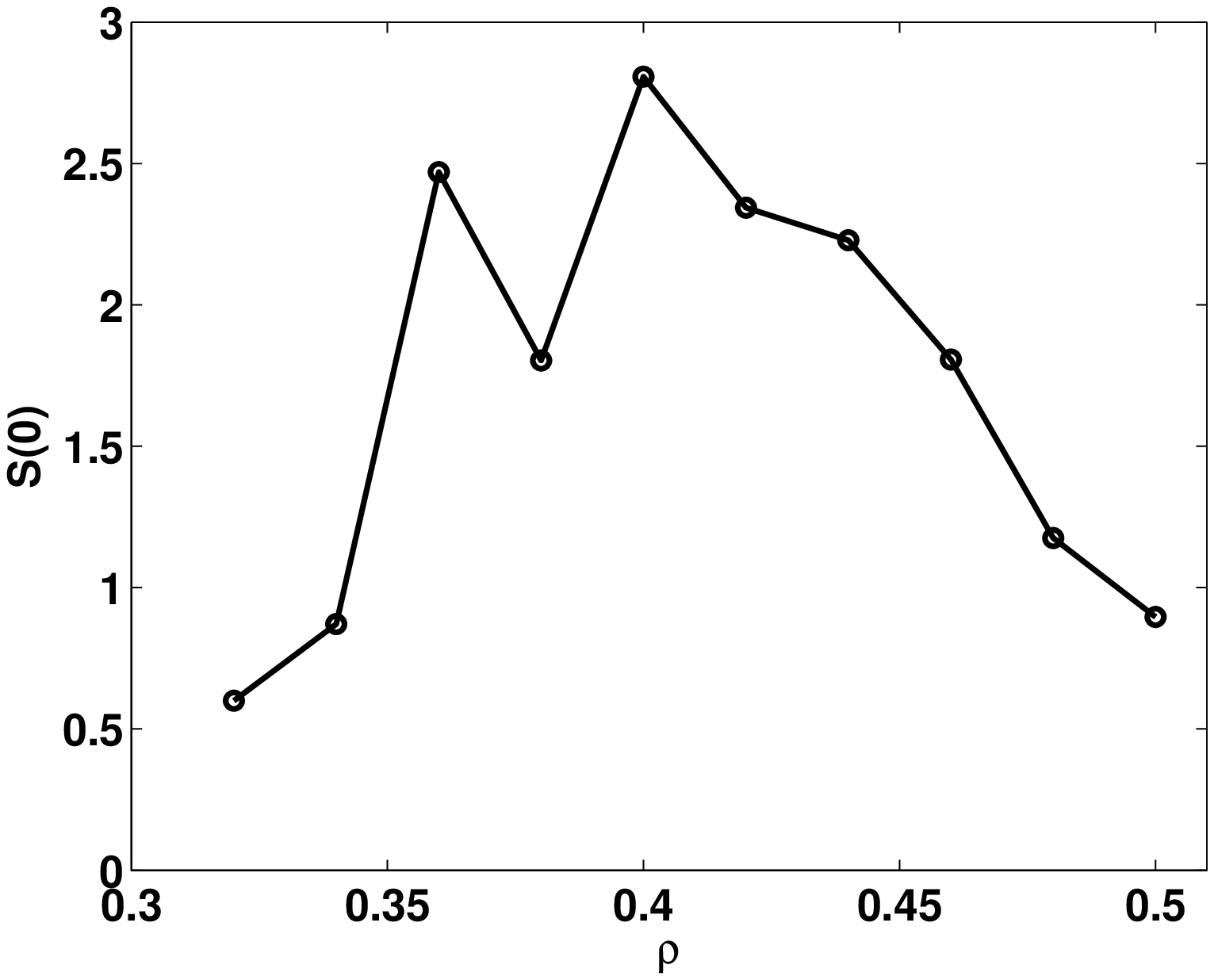}
\end{figure}

1d

\newpage
\begin{figure}
  \includegraphics[width=10.cm]{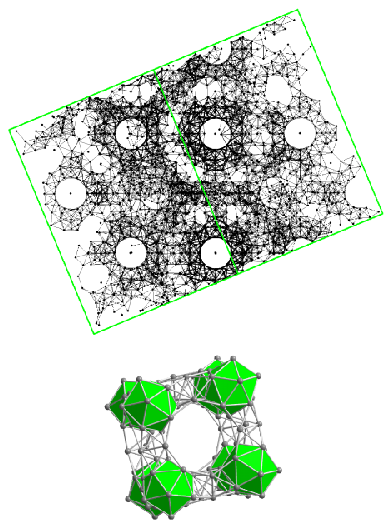}
\end{figure}

2a

\newpage
\begin{figure}
  \includegraphics[width=9.cm]{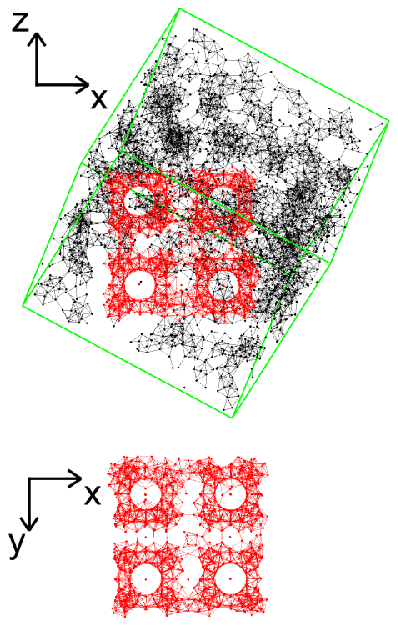}
\end{figure}

2b

\newpage
\begin{figure}
  \includegraphics[width=10.cm]{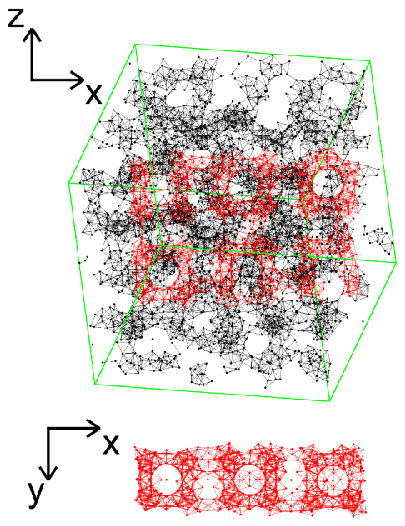}
\end{figure}

2c

\newpage
\begin{figure}
  \includegraphics[width=10.cm]{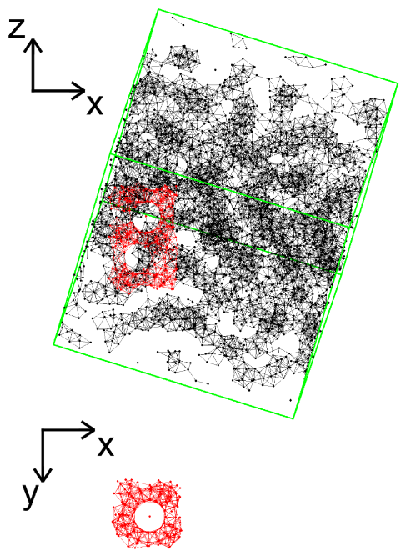}
\end{figure}

2d

\newpage
\begin{figure}
  \includegraphics[width=10.cm]{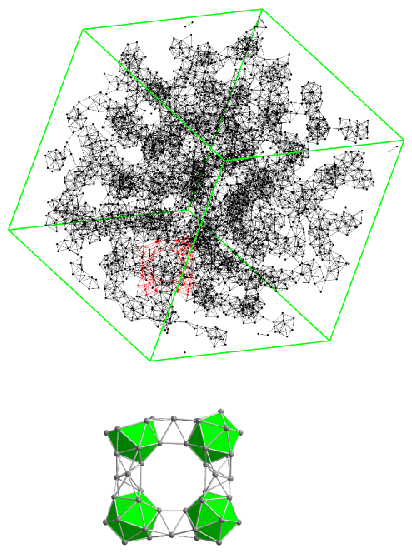}
\end{figure}

2e

\newpage
\begin{figure}
  \includegraphics[width=16.cm]{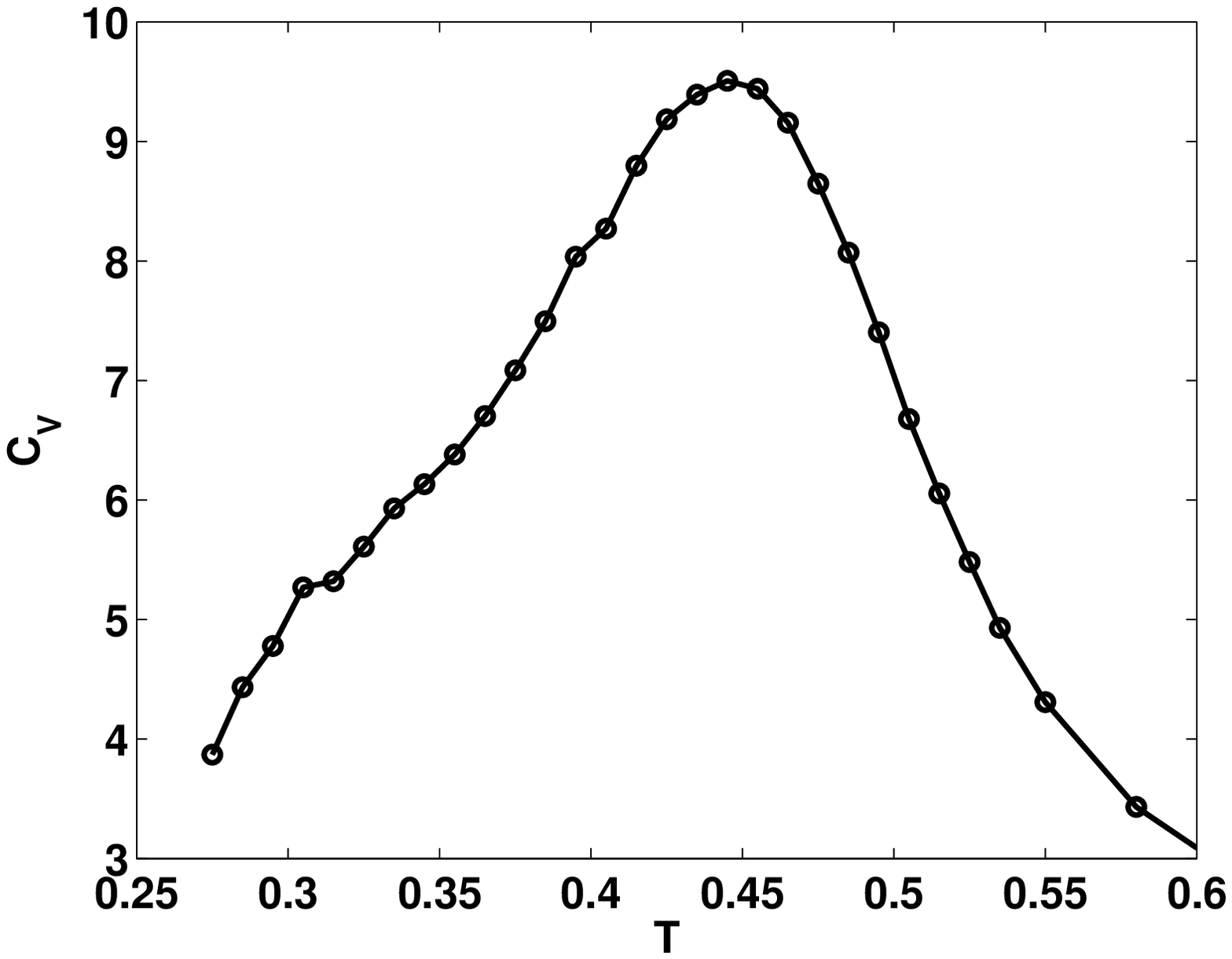}
\end{figure}

3a

\newpage
\begin{figure}
  \includegraphics[width=16.cm]{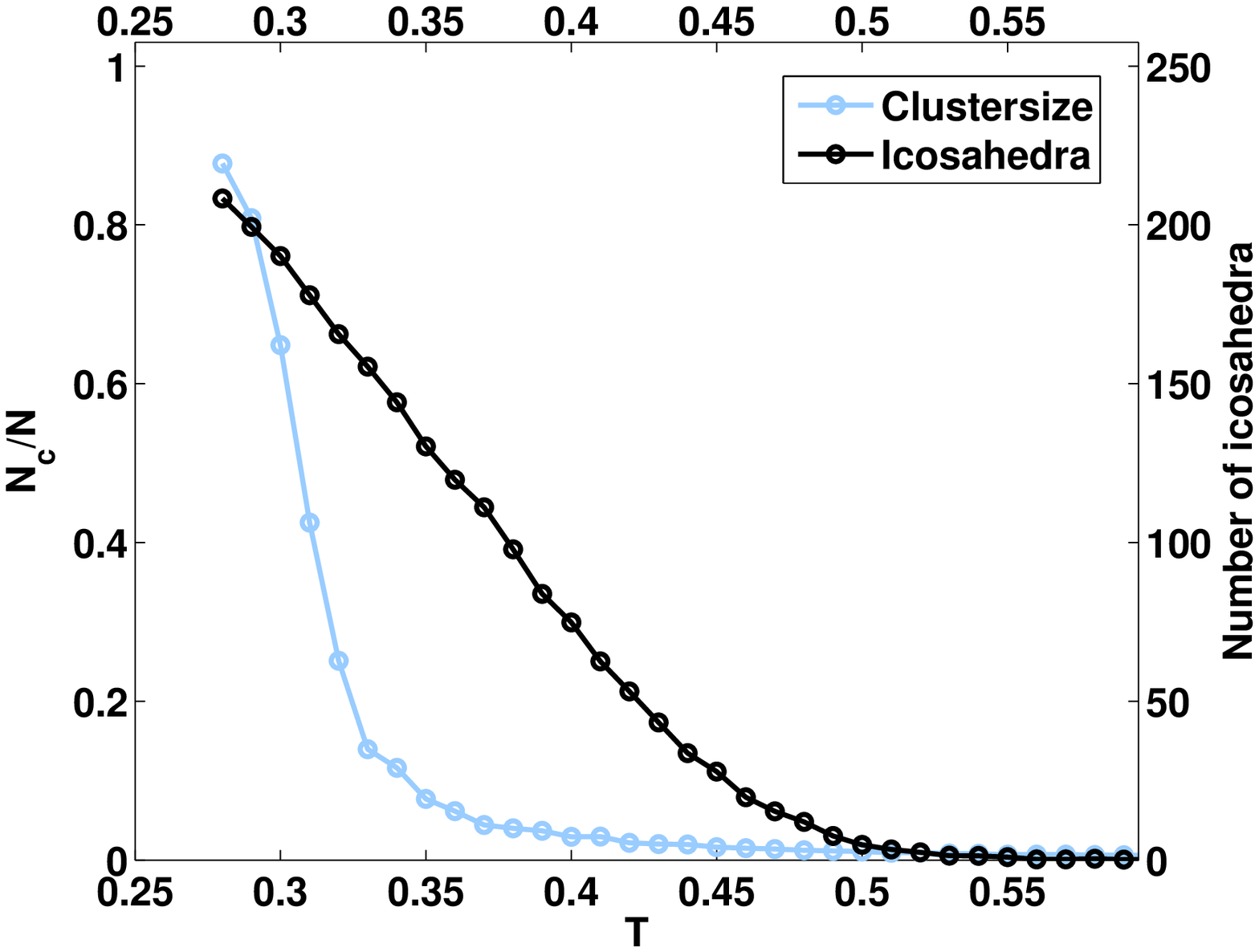}
\end{figure}

3b

\newpage
\begin{figure}
  \includegraphics[width=16.cm]{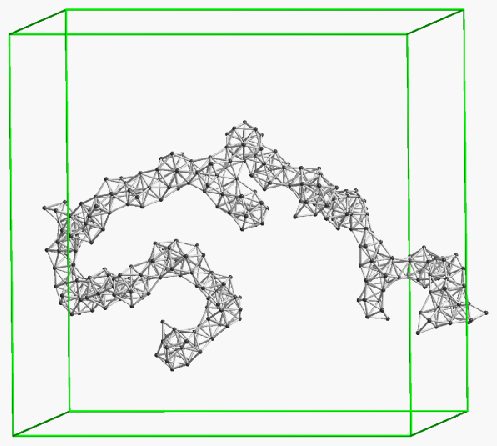}
\end{figure}

3c

\newpage
\begin{figure}
  \includegraphics[width=16.cm]{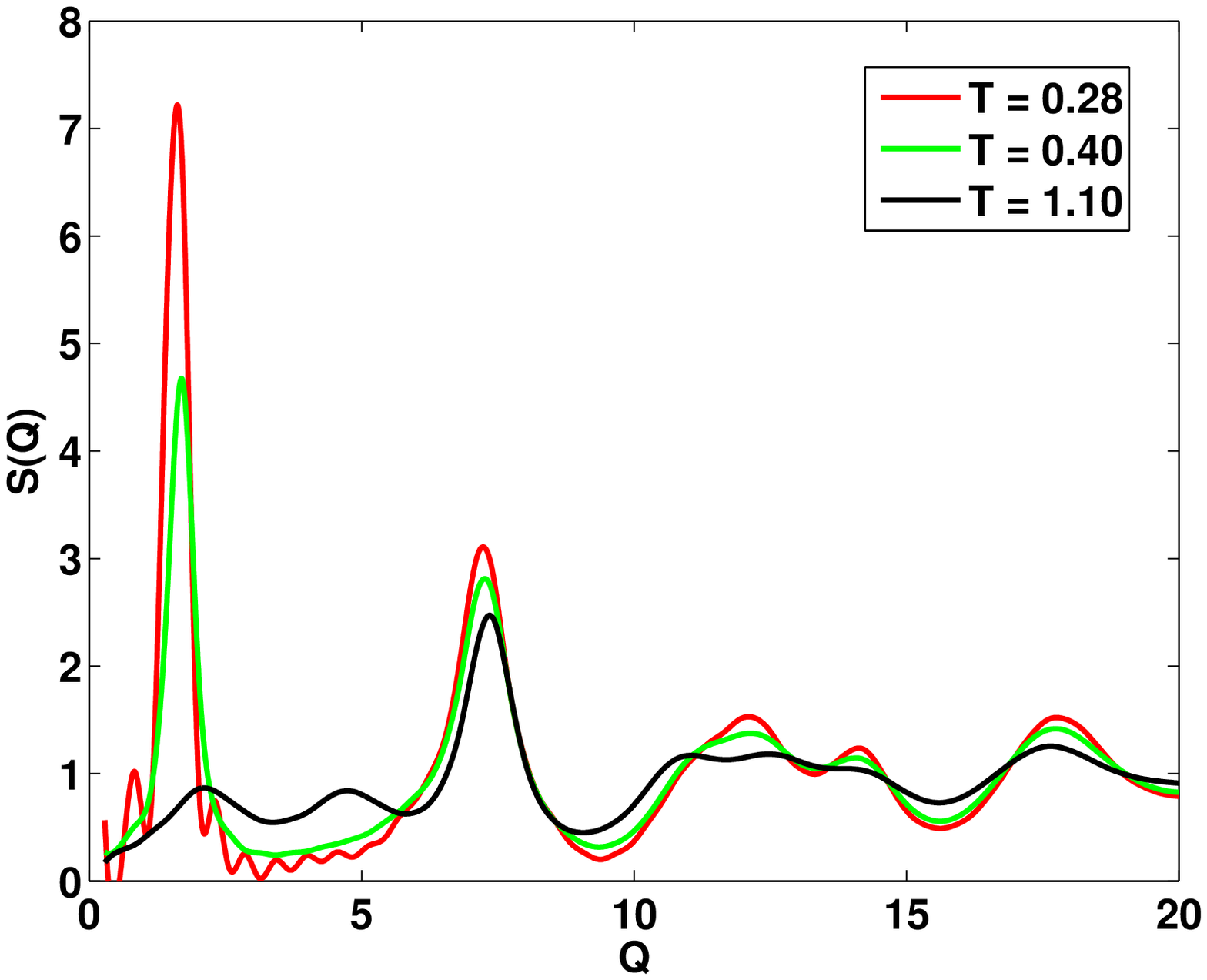}
\end{figure}

3d

\newpage
\begin{figure}
  \includegraphics[width=16.cm]{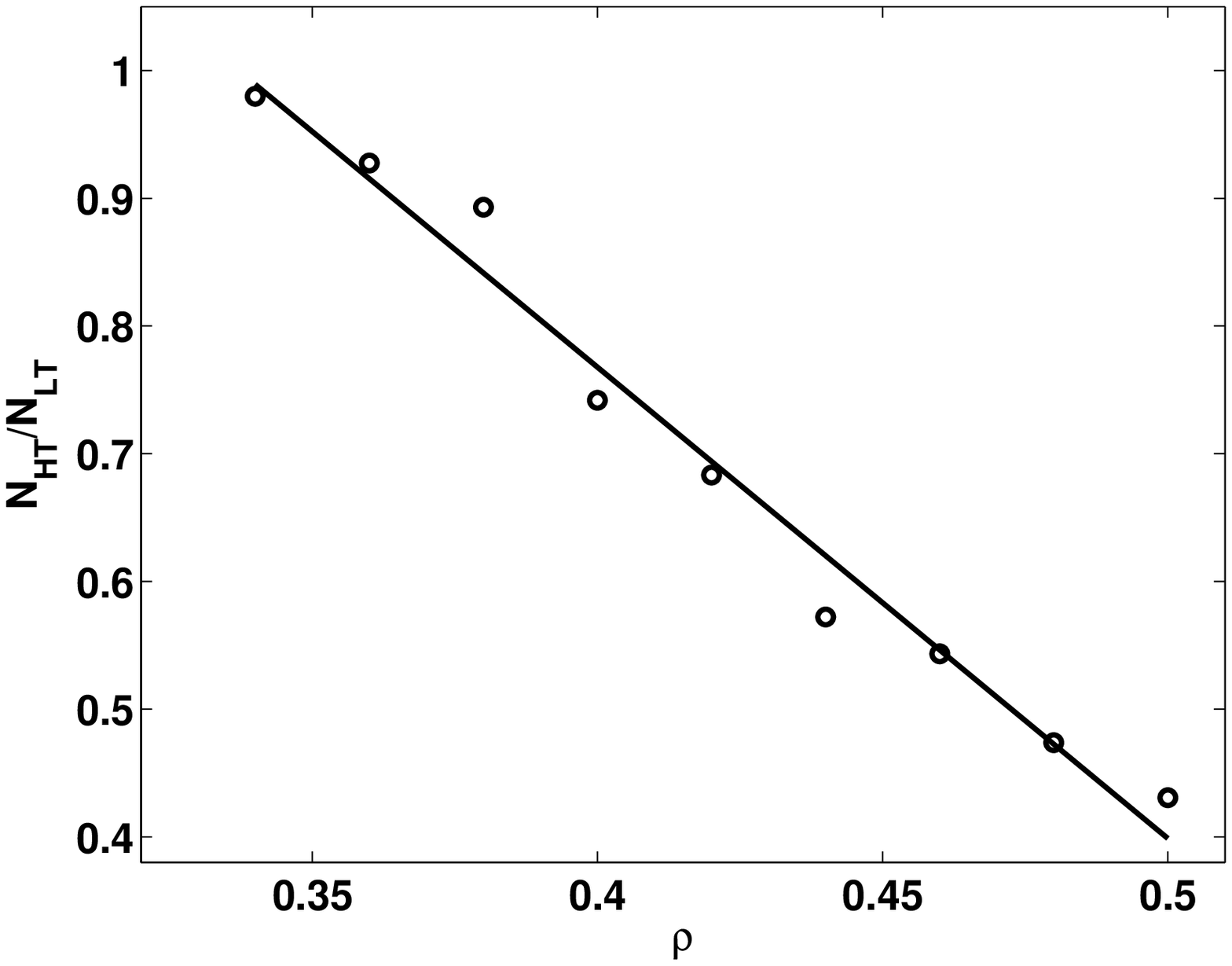}
\end{figure}

3e

\newpage
\begin{figure}
  \includegraphics[width=16.cm]{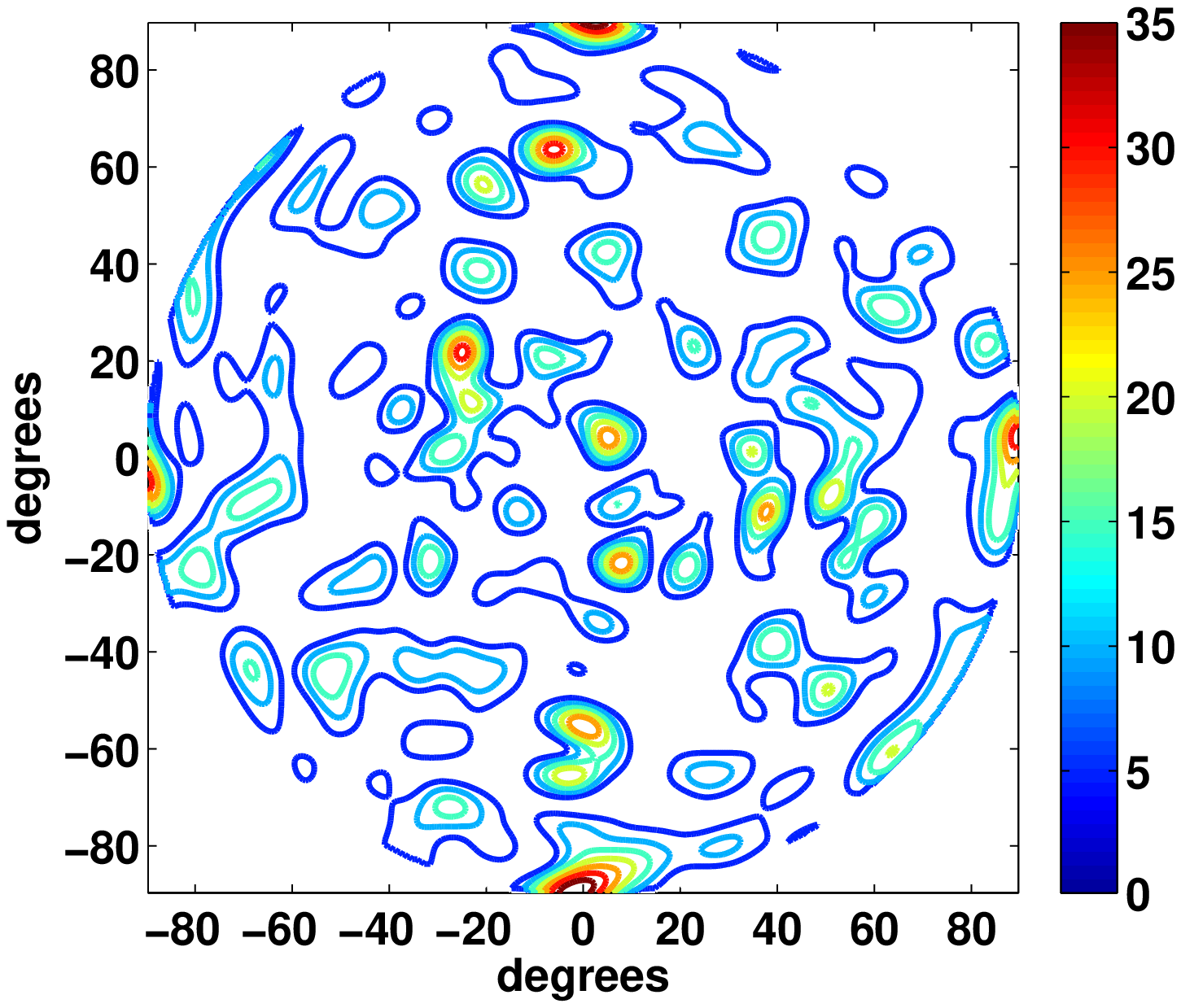}
\end{figure}

4a

\newpage
\begin{figure}
  \includegraphics[width=16.cm]{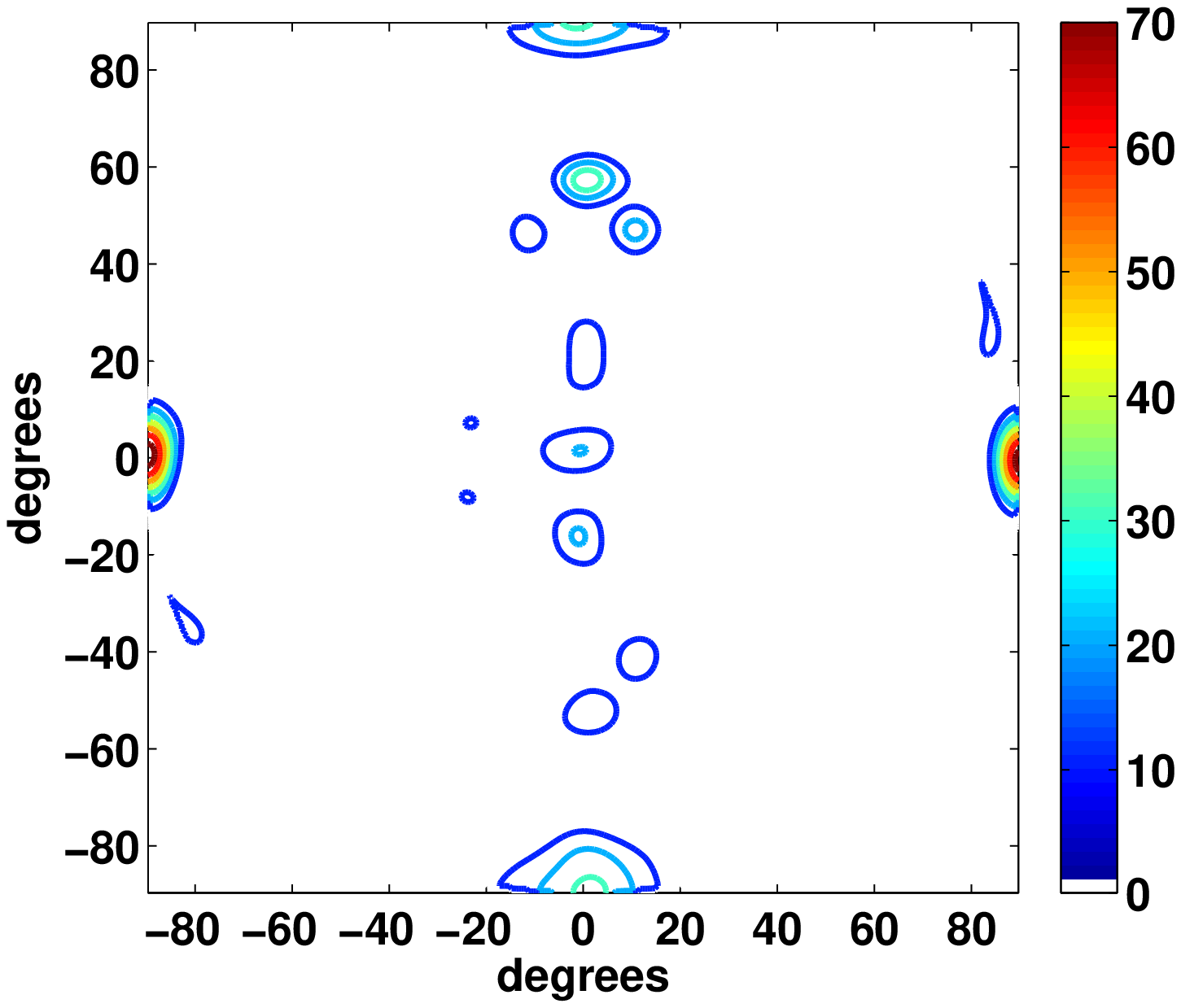}
\end{figure}

4b

\newpage
\begin{figure}
  \includegraphics[width=16.cm]{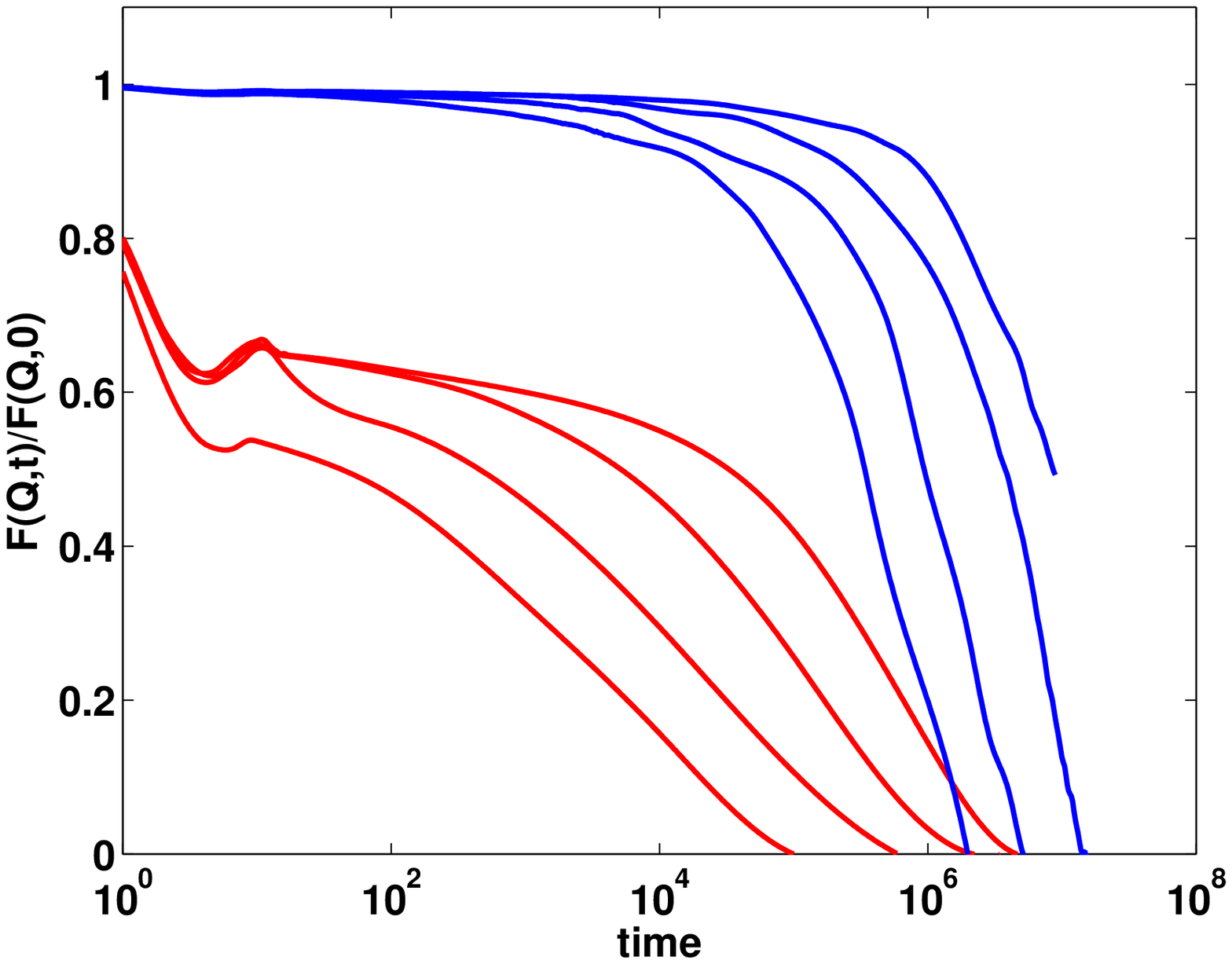}
\end{figure}

4c

\end{document}